\documentclass[twocolumn,showpacs,amsmath,amssymb,amsfonts,floatfix,prl,aps]{revtex4}
\usepackage{graphicx}
\usepackage{color}
\pdfoutput=1

\begin{document}
\title{Bootstrap approximation for the exchange-correlation kernel of time-dependent density
 functional theory}
\author{S. Sharma}
\email{sharma@mpi-halle.mpg.de}
\author{J. K. Dewhurst}
\author{A. Sanna}
\author{E. K. U. Gross}
\affiliation{Max-Planck-Institut f\"ur Mikrostrukturphysik, Weinberg 2, 
D-06120 Halle, Germany.}
\date{\today}

\begin{abstract}
A new parameter-free approximation for the exchange-correlation kernel $f_{\rm xc}$ of time-dependent density
functional theory is proposed. This kernel is expressed as an algorithm in which the exact
Dyson equation for the response as well as a further approximate condition are solved
together self-consistently leading to a simple parameter-free kernel.
We apply this to the calculation of optical spectra for various small bandgap
(Ge, Si, GaAs, AlN, TiO$_2$, SiC), large bandgap 
(C, LiF, Ar, Ne) and magnetic (NiO) insulators. The calculated spectra are in very good
agreement with experiment for this diverse set of materials, highlighting the universal
applicability of the new kernel.
\end{abstract}

\maketitle

%%%%%%%%%%%%%%
% Introduction
%%%%%%%%%%%%%%
The \emph{ab-initio} calculation of optical absorption spectra of nano-structures 
and solids is a formidable task. The current state-of-the-art is based on many-body 
perturbation theory. A typical calculation involves two distinct steps:
First, the quasi-particle spectral density function is calculated using the $GW$
approximation, yielding accurate electron removal and addition energies, and therefore 
a good prediction for the fundamental gap\cite{gw}. In the second step, the 
Bethe-Salpeter equation (BSE) is solved using the one-body Green's function obtained in 
the $GW$ step. Resonances, corresponding to bound electron-hole pairs called excitons,
which have energies inside the gap, can then appear in the spectrum.
The two step procedure described above is a well-established method for yielding 
macroscopic dielectric tensors which are generally in good agreement with 
experiment\cite{onida02,onida95,albrecht98,benedict98,rohlfing98,marini09}.
Unfortunately, solving the BSE involves diagonalizing a large matrix which couples different
Bloch state $k$-points. As a consequence, the method is computationally expensive.

Time-dependent density functional theory (TDDFT)\cite{tddft}, which
extends density functional theory into the time domain, is another method
able, in principle, to determine neutral excitations of a system. 
Although formally exact, the predictions of TDDFT are only as good as the approximation of
the exchange-correlation (xc) kernel:
$f_{\rm xc}({\bf r},{\bf r}',t-t')\equiv\delta v_{\rm xc}({\bf r},t)/\delta\rho({\bf r}',t')$,
where $v_{\rm xc}$ is the TD exchange-correlation potential and $\rho$ is the TD density.
There are several such approximate kernels in existence, the earliest of which
is the adiabatic local density approximation (ALDA)\cite{alda}, where $v_{\rm xc}({\bf r},t)$
is determined from the usual ground-state local density approximation (LDA),
calculated instantaneously for $\rho({\bf r},t)$. In practice however,
the macroscopic dielectric function calculated using this kernel has two well-known 
deficiencies: the quasi-particle gap is too small and the physics of the bound electron-hole 
pair is totally missing -- in fact ALDA does not improve on the results obtained within the 
random phase approximation (RPA) which corresponds to the trivial kernel 
$f_{\rm xc}=0$\cite{gavri97}. In the present work we concentrate on the second of these 
problems, namely the missing excitonic peak in the spectrum. There have been previous 
attempts to solve this problem\cite{botti07}, and there exist kernels which correctly 
reproduce the peaks in the optical spectrum associated with bound excitons. The nano-quanta 
kernel by Sottile {\it et al.}\cite{sot03}, derived from the four-point Bethe-Salpeter kernel,
is very accurate but has the drawback of being nearly as computationally demanding as 
solving the BSE itself. The long-range correction (LRC) kernel\cite{reining02,botti05} 
has a particularly simple form in reciprocal space, $f_{\rm xc}=-\alpha/q^2$, which limits its
computational cost. This kernel produces the desired excitonic peak, but depends on the choice 
of the parameter $\alpha$, which turns out to be strongly material-dependent, thereby limiting 
the predictiveness of this approximation. In the present work we propose a new parameter-free 
approximation for $f_{\rm xc}$, and demonstrate that this kernel is as accurate as BSE with a
computational cost of ALDA.

%%%%%%%%%%%%%%
% functional
%%%%%%%%%%%%%%
The exact relationship between the dielectric function
$\varepsilon$ and the kernel $f_{\rm xc}$ for a periodic solid can be written as
\begin{align}\label{dyson}
 \varepsilon^{-1}({\bf q},\omega)&= 1+v({\bf q})\chi({\bf q},\omega)\nonumber\\
 &=1+\frac{v({\bf q})\chi_0({\bf q},\omega)}
 {1-\left[v({\bf q})+f_{\rm xc}({\bf q},\omega)\right]\chi_0({\bf q},\omega)},
\end{align}
where $v$ is the bare Coulomb potential, $\chi$ is the full response function,
and $\chi_0$ is the response function of the non-interacting Kohn-Sham system.
All these quantities are matrices in the basis of reciprocal lattice vectors
${\bf G}$.
The bootstrap kernel is a frequency-independent approximation given by:
\begin{align}\label{app}
 f^{\rm BS}_{\rm xc}({\bf q},\omega)=
 -\frac{\varepsilon^{-1}({\bf q},\omega=0)v({\bf q})}
 {\varepsilon_0({\bf q},\omega=0)-1}
\end{align}
where $\varepsilon_0({\bf q},\omega)\equiv 1-v({\bf q})\chi_0({\bf q},\omega)$. 
$\varepsilon^{-1}({\bf q},\omega=0)$ is determined \emph{self-consistently} with 
Eq. (\ref{dyson}). 
We note that although Eq. (\ref{dyson}) is exact, it is useful only when either
$f_{\rm xc}$ or $\varepsilon$ is given; if neither are available then obviously 
it cannot be used as a generating equation for both quantities. With the
addition of the approximation given by Eq. (\ref{app}) however, both $f_{\rm xc}$ 
and $\varepsilon$ can be determined from knowledge of $\chi_0$ exclusively. The 
{\it modus operandi} for doing so, is to start by setting $f_{\rm xc}=0$ and then 
solving Eq. ({\ref{dyson}) to obtain $\varepsilon^{-1}$. This is then used in 
Eq. (\ref{app}) to find a new $f_{\rm xc}$, and the procedure repeated until 
self-consistency between the two equations is achieved. The advantages of this form 
for the kernel is that 1. it has the correct 1/$q^2$ behavior\cite{onida02,ghosez97}; 
2. as $\varepsilon$ improves from cycle to cycle so does $f_{\rm xc}$; 3. the computation 
cost is minimal, as the most expensive part is the calculation of $\chi_0$, which needs
to calculated only once; and 4. most importantly, no system-dependent external
parameter is required.

The $\chi_0$ in Eqs. (\ref{dyson}) and (\ref{app}) are, in practice, calculated using an 
approximate ground state xc functional, such as the LDA. To overcome the shortcomings of 
such an approximation, we further replace the $\chi_0$ by a model response function $\chi_m$ 
coming either from scissors-corrected LDA, or from $GW$ or from LDA+$U$. This has the 
advantage that $\chi_m$, and consequently $\chi$, has the correct gap to begin with. From 
the formal point of view, this replacement amounts to approximating the TDDFT kernel by 
\begin{align}\label{H1}
f^{\rm appr}_{\rm xc}({\bf q},\omega)= \frac{1}{\chi_0({\bf q},\omega)}-
\frac{1}{\chi_m({\bf q},\omega)}+f^{\rm BS}_{\rm xc}({\bf q})
\end{align}
%%%%%%%%%%%%%%
% ELK
%%%%%%%%%%%%%%
Using the method outlined above, optical spectra for various extended systems\cite{latp} were
calculated using the full-potential linearized augmented plane wave (FP-LAPW)
method \cite{Singh}, implemented within the Elk code \cite{elk}.
Except for the case of solid Ar, a shifted $k$-point mesh of $15\times15\times15$ is
used to ensure convergence\cite{smear}. In the case of solid Ar a shifted mesh of
$25\times25\times25$ $k$-points was
required for convergence of the optical spectrum. All the calculations
were performed by scissor shifting the ground-state Kohn-Sham eigenvalues to
reproduce the experimental bandgap.

%%%%%%%%%%%%%%
% Results
%%%%%%%%%%%%%%
\begin{figure}[ht]
\centerline{\includegraphics[width=\columnwidth,angle=-0]{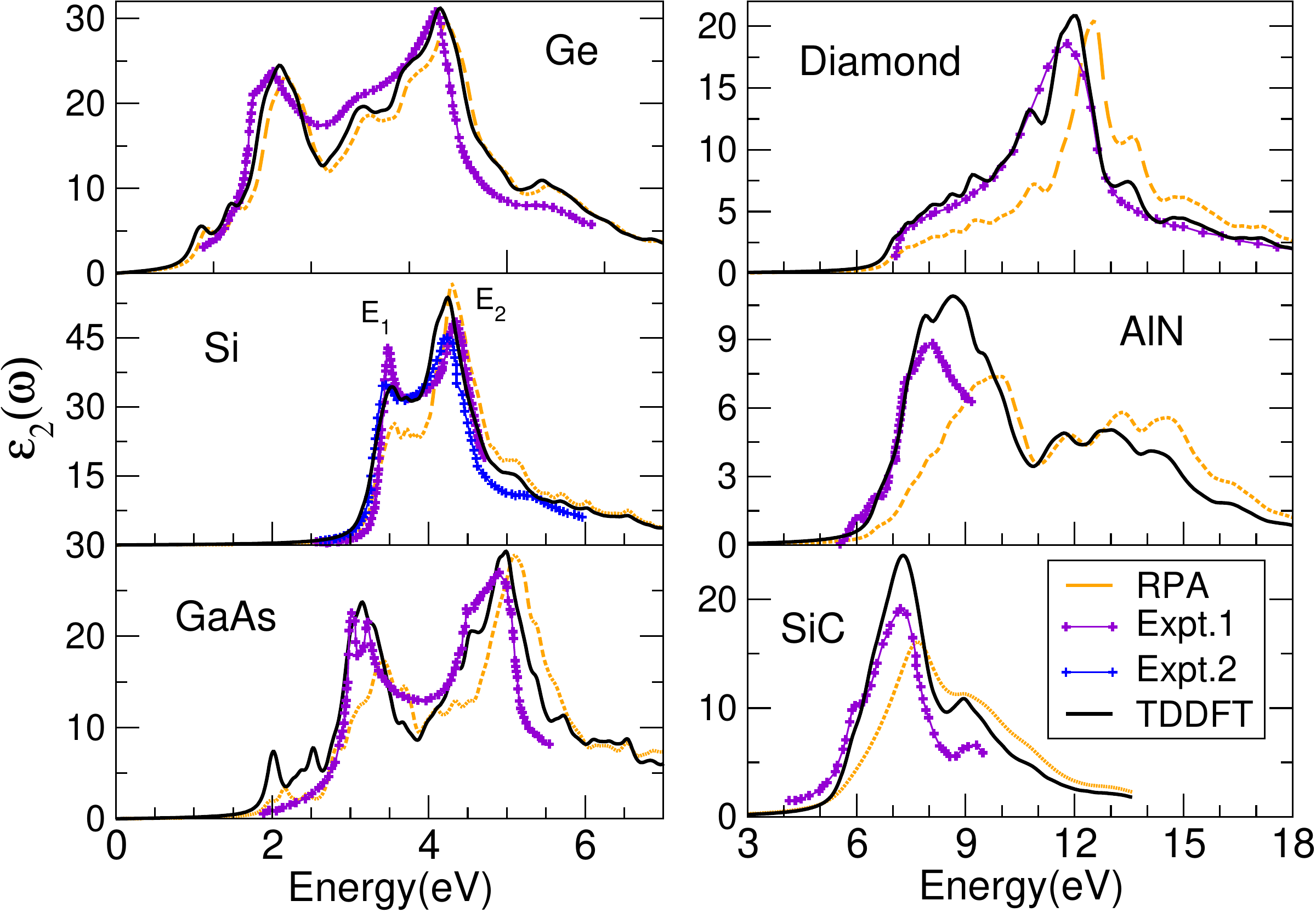}}
\caption{(Color online)  Imaginary part of the dielectric tensor ($\varepsilon_2$)
as function\cite{smear} of photon energy (in eV). Experimental data 
are taken from the following sources:
Ge from \onlinecite{ge-exp}, Si from \onlinecite{si-exp} and \onlinecite{si-exp2}, 
GaAs from \onlinecite{gaas-exp}, diamond from \onlinecite{diam-exp}, 
AlN from \onlinecite{aln-exp}, and SiC from \onlinecite{sic-exp}}
\label{set1}
\end{figure}

Presented in Fig. \ref{set1} are the results for some small (Ge $\sim 0.67$ eV)
to medium (diamond $\sim 5.47$ eV) bandgap semiconductors.
For comparison, experimental data as well as the RPA spectra are also plotted. 
The experimental data clearly show that all these materials have weakly bound
excitons leading to a small shifting of the spectral 
weight to lower energies compared to RPA. 
The results from TDDFT with the new kernel exactly follow this trend and
are in excellent overall agreement with experiment.

%Ge
For Ge the TDDFT results are only slightly
different from the RPA values which themselves are in agreement with experiment.
It is clear that for Ge the RPA is enough and $f_{\rm xc}$ does not significantly
improve on the result.
%Si
This is in complete contrast to the spectrum of Si  where
the spectral 
weight is redistributed and, corresponding to experiment, the TDDFT results show
an enhanced E$_1$ peak. 
The height of the E$_2$ peak remains marginally overestimated by TDDFT.
This overestimation
is not particular to the present approximation for $f_{\rm xc}$, it is also a
feature of the BSE-derived kernel \cite{sot03}.
%GaAs
The dielectric function for GaAs is also in very good
agreement with experiment -- subtle features like
the kink at 4.25 eV is well captured by our bootstrap procedure. 

The second column of Fig. \ref{set1} contains results for medium bandgap
insulators. In all these materials a
significant redistribution of the spectral weight to lower photon energies is observed. 
%C
For diamond, the bootstrap procedure  correctly leads to an
enhancement of the shoulder at low photon energies.
The position of the main peak around 12 eV is shifted to lower energies, and
the whole spectrum is in near perfect agreement with experiment. 
%AlN
AlN is a paticularly interesting
case, TDDFT  shifts the spectral weight to lower energies,
and although the height of the peak is too large, the agreement with experiment
is considerably better than that obtained by the equivalent BSE
calculation \cite{bechstedt05,hahn05,lask07}.
%SiC
For SiC the results show an improvement over the RPA
spectrum, but the height of the main peak as well 
as the shoulder at 9 eV are overestimated. This trend is also observed in
previous BSE results\cite{botti04-1}.

\begin{figure}[ht]
\centerline{\includegraphics[width=\columnwidth,angle=-0]{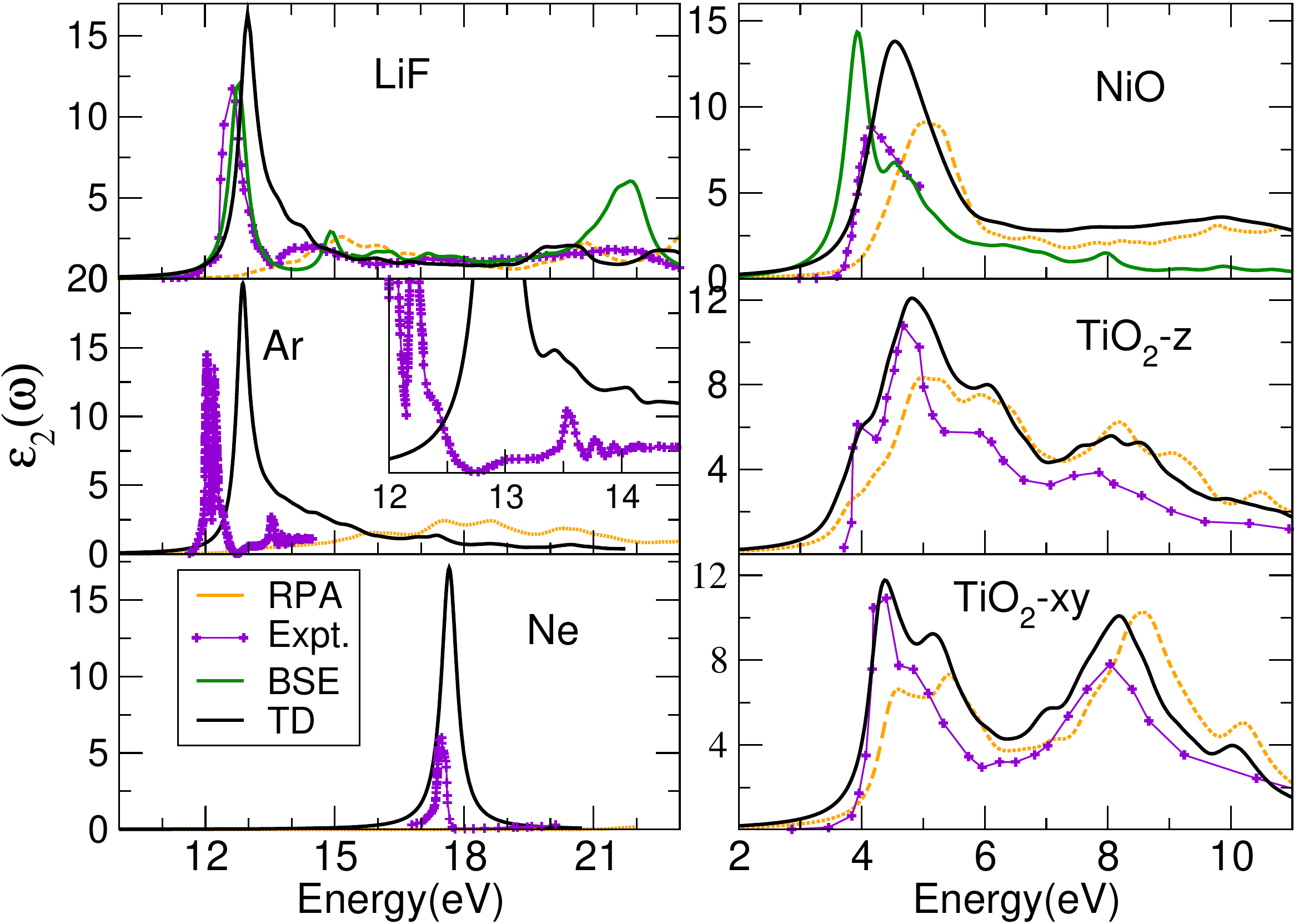}}
\caption{(Color online) Imaginary part of the dielectric tensor ($\varepsilon_2$)
as function\cite{smear} of photon energy (in eV). Experimental data are taken
from the following sources: LiF from \onlinecite{lif-exp}, Ar and Ne from \onlinecite{arne-exp},
NiO from \onlinecite{nio1-exp} and TiO$_2$ from \onlinecite{tio2-exp}. In the inset a smaller 
broadening is used to resolve the peaks better\cite{smear}}
\label{set2}
\end{figure}

A stringent test for any approximate xc-kernel is in its ability to treat
materials with strongly bound excitons. In these cases a new resonant peak
appears in the bandgap itself and represents the bound state of an electron-hole
pair. Perhaps the most studied test case for this phenomenon is the ionic solid LiF.
Other excitonic materials which have also attracted attention and are considered
particularly difficult to treat are the 
noble gas solids. Plotted in the first column of Fig. \ref{set2} are the
results for three materials of this class: LiF, solid Ar and Ne.
What is immediately clear is that the bootstrap procedure, which gave only a
slight shift of spectral weight for
Ge, now gives rise to an entirely new bound excitonic peak inside the
gap in all three cases. The location of the peak, which corresponds to the
excitonic binding energy, is also very well reproduced for all these materials.

%LiF
Despite a good overall agreement we find that for LiF the main peak at 12.5 eV is overestimated, 
and the peak at 14.3 eV appears as a hump in the TDDFT results. 
Nevertheless, it is encouraging to note that the BSE spectrum, as well that obtained using the BSE-derived
kernel \cite{marini03}, include a spurious peak 
at around 21 eV which is absent in the present calculations. 
%Ar
Noble gas solids have very weak band dispersion and polarazibility, which results
in very strongly bound electron-hole pairs.
In the case of solid Ar, one can observe a strongly localised Frenkel exciton\cite{sot07} at about
12 eV and a Wannier exciton at about 14 eV.
This physics is totally missing within the RPA. Remarkably though,
the bootstrap procedure captures both these excitons, although the Wannier
exciton is suppressed (see inset). Exactly like in BSE and LRC calculations\cite{sot07}, the 
Frenkel exciton is under-bound by 0.7eV.
%Ne
Ne has a strongly bound Frenkel exciton and the present calculations capture the
corresponding excitonic peak. Similar to the BSE results\cite{mula05}, the height of
this peak is overestimated by our TDDFT calculations.

%NiO
The second column of Fig. \ref{set2} consists of some special cases -- NiO has
an anti-ferromagnetic ground state,
and the LDA+$U$ method is needed to obtain a physically reasonable band structure
for this material. This material provides the bootstrap technique
with a test of its validity for magnetic materials and also with a check of its
performance when the scissors-corrected LDA is replaced by LDA+$U$, where $U$ is chosen
to reproduce the experimental gap. It is clear from  Fig. \ref{set2} that the bootstrap
method leads once again to the correct excitonic binding energy. The experimental data
for NiO are rather old and substantially broadened \cite{nio1-exp},
and, assuming the veracity of these data, both TDDFT and BSE\cite{rodl-thesis} overestimate
the peak height. It is worth noting that the BSE spectrum is redshifted
relative to experiment, while
the TDDFT spectrum is correctly positioned.
% TiO2
Results for the anatase phase of TiO$_2$ are also presented in Fig. \ref{set2}.
This material is important for its industrial
use in photovoltaics and has been well characterized using the BSE and GW
method \cite{lawler08,kang10} as well as experiment\cite{tio2-exp}. TiO$_2$ is a
useful test for the bootstrap method due to its non-cubic unit cell, which leads
to directional anisotropy in the optical spectrum. As can be seen in
Fig. \ref{set2}, the bootstrap method captures this anisotropy very well indeed.
Even subtle features like the small shoulder at
$\sim 4$ eV in the out-of-plane dielectric function, which is missing in the
in-plane case, are well reproduced.
We find that our peak heights are slightly overestimated, which is also the
case for the BSE results \cite{lawler08}. 

%%%%%%%%%%%%%%%%%
% Real part
%%%%%%%%%%%%%%%%%
\begin{figure}[ht]
\centerline{\includegraphics[width=\columnwidth,angle=-0]{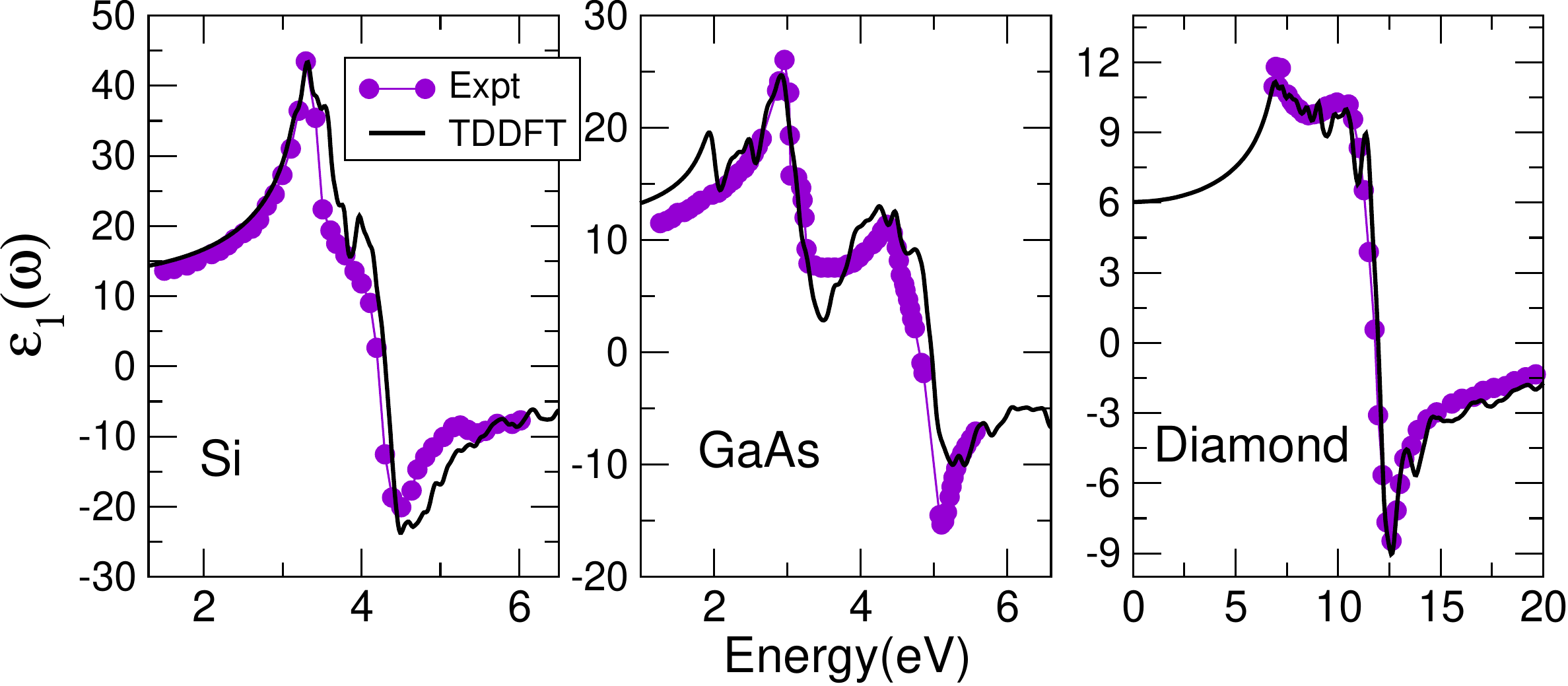}}
\caption{(Color online) Real part of the dielectric tensor ($\varepsilon_1$)
as function of photon energy (in eV) for Si, GaAs and diamond.}
\label{set3}
\end{figure}
It is also interesting to compare the real part of the dielectric function with
available experimental data. Results for Si, GaAs and diamond are presented in
Fig. \ref{set3}. In all three cases, TDDFT results are in excellent agreement with
the experimental data. We note that in the low frequency regime (below 2eV) the
TDDFT results for GaAs deviate from experiments in exactly the same manner as
found for the LRC kernel\cite{botti04-1}. 

%%%%%%%%%%%%%%%%%
% convergence
%%%%%%%%%%%%%%%%%
\begin{figure}[ht]
\centerline{\includegraphics[width=\columnwidth,angle=-0]{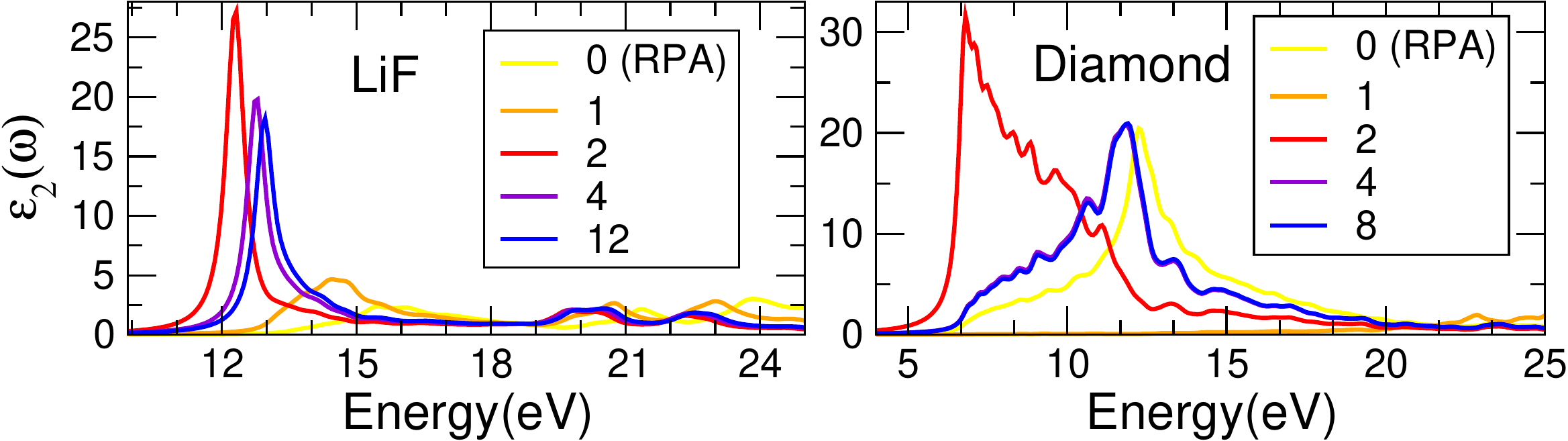}}
\caption{(Color online) Imaginary part of the dielectric tensor ($\varepsilon_2$)
as function of photon energy (in eV) for LiF and diamond. $\varepsilon_2$ is plotted 
at various steps of the self-consistency.}
\label{conv}
\end{figure}
The effect of the bootstrap procedure can be seen in Fig. \ref{conv}
where  $\varepsilon_2$ is potted at each step of the self-consistency.  
As specified, the starting point is $f_{\rm xc}=0$ which yields the RPA spectrum in 
the first iteration. From then on, $\varepsilon_2$ changes 
substantially until convergence, indicating the importance of the self-consistency.

%%%%%%%%%%%%%%%%%
% draw parallels
%%%%%%%%%%%%%%%%%
An interesting point to note about Eq. (\ref{app}) is that this form of $f_{\rm xc}$ 
is related to the two existing TDDFT kernels which capture the excitonic physics.
In particular it corresponds to the LRC kernel when
$\alpha=4\pi\varepsilon^{-1}/(\varepsilon_0-1)$,
and $\varepsilon$ is determined self-consistently.
Also, like for the LRC kernel, the ${\bf G}={\bf G}'=0$ part of 
$f_{\rm xc}$ is by far the most important contribution.
This also explains the fast convergence of the optical spectra 
with respect to the reciprocal lattice vectors included in $f_{\rm xc}$.
Comparing our kernel to the BSE-derived approximation, in both cases
the $f_{\rm xc}$ is
proportional to the screened Coulomb matrix
elements $\varepsilon^{-1}v$. Choosing $f_{\rm xc}$ to be 
proportional to the screened Coulomb interaction was also exploited by Turkowski {\it et al.}
(see Eq. 1 and 15 of Ref. \onlinecite{turk09}).
There remain several interesting aspects of the bootstrap method to be explored in the future:
1. the finite ${\bf q}$ version of this approximation to determine the energy
loss spectrum; 2. performance of this approximation for two-dimensional systems,
like graphene sheets or even nanotubes where
excitonic effects are particularly strong; 
3. to study the possibility of making this functional frequency-dependent; and
4. to explore analogous construction of $f_{\rm xc}$ for finite systems and study its
performance for excitonic spectra in molecular aggregates.

%%%%%%%%%%%%%%
% conclusions
%%%%%%%%%%%%%%
Thus we have demonstrated that the bootstrap procedure gives a parameter-free
TDDFT kernel which yields very accurate optical spectra. 
The same functional which produces a small shift, relative to the RPA, in
the absorption edge of Ge,
also generates an entirely new excitonic peak within the bandgap of LiF, Ar and Ne.
This indicates that the bootstrap kernel has wide applicability.
The same quality of results can also be obtained by solving the BSE,
but the present method is highly desirable from the point of view of
computation effort.

%\bibliography{fxc}

\end{document}